\documentclass[reprint,superscriptaddress,nofootinbib,notitlepage,nobibnotes,amsmath,amssymb,aps,pre,floatfix,raggedbottom]{revtex4-2}
\usepackage{graphicx}
\usepackage{amsmath,amssymb,amsfonts,amsthm,bm} 
\usepackage{hyperref}
\usepackage[version=4]{mhchem} 
\usepackage[separate-uncertainty=true,multi-part-units=single]{siunitx}

\newcommand{\rb}[1]{#1}

\newcommand{\avg}[1]{\langle#1\rangle}
\DeclareMathOperator{\const}{\mathrm{const}}
\DeclareMathOperator{\me}{\mathrm{e}}

\begin{document}
\title{Statistical physics of inhomogeneous transport: \texorpdfstring{\\}{} Unification of diffusion laws and inference from first-passage statistics}
\date{\today}
\author{Roman Belousov}\email{belousov.roman@gmail.com}
\author{Ali Hassanali}\email{ahassana@ictp.it}
\author{\'Edgar Rold\'{a}n}\email{edgar@ictp.it}
\affiliation{
  ICTP - The Abdus Salam International Centre for Theoretical Physics, Strada Costiera 11, 34151 Trieste, Italy
}

\begin{abstract}
  Characterization of composite materials, whose properties vary in space over microscopic scales, has become a problem of broad interdisciplinary interest. In particular, estimation of the inhomogeneous transport coefficients, e.g. the diffusion coefficient or the heat conductivity which shape important processes in biology and engineering, is a challenging task. The analysis of such systems is further complicated, because two alternative formulations of the inhomogeneous transport equations exist in the literature---the Smoluchowski and Fokker-Planck equations, which are also related to the so-called Ito-Stratonovich dilemma. Using the theory of statistical physics, we show that the two formulations, usually regarded as distinct models, are physically equivalent. From this result we develop efficient estimates for the transverse space-dependent diffusion coefficient in fluids near a phase boundary. Our method requires only measurements of escape probabilities and mean exit times of molecules leaving a narrow spatial region. We test our estimates in three case studies: (i) a Langevin model of a B\"uttikker-Landauer ratchet;  atomistic molecular-dynamics simulations  of liquid-water molecules in contact with (ii) vapor and (iii) soap (surfactant) film which has promising applications in physical chemistry. Our analysis reveals that near the surfactant monolayer the mobility of water molecules is slowed down almost twice with respect to the bulk liquid. Moreover, the diffusion coefficient of water correlates with the transition from hydrophilic to hydrophobic parts of the film.
\end{abstract}
\maketitle

\section{Introduction}
Various physical processes in nature and engineering are shaped by composite materials
with inhomogeneous transport properties, e.g. space-dependent mass diffusion and heat conductivity~\rb{\cite{Hanggi1980connection,klemens1989,knapp1993,muegge1995,zabolotsky1993,dramicanin1998,divo2000,kruse2004dpp,zhang2004,ao2007existence,godec2009,best2010,hinczewski2010,rajabpour2011,celani2012anomalous,hoang2012,dentzDiffusionTrappingHeterogeneous2012,best2013,chen2015,jabbari2017,yang2017,nikonenko2018,fusi2019,zou2019,belousov2020,liu2004,liu2005,falciani2020,richard2019active,willems2020,richardActiveCargoPositioning2019,xu2020heterogeneous,villarruelHighRatesCalciumfree2021,corti2021structure,Rondoni2021,devosMembranePermeabilityCharacteristic2018,boStochasticDynamics2021}}. Modelling of such systems
subsists the quantitative description of living matter and guides the design of nanoscale devices.
In this line of research, estimation of the physical  properties of inhomogeneous systems
 constitutes a  problem of broad interdisciplinary interest.

Curiously, inhomogeneous transport problems have created a controversy between two
theoretical approaches in statistical mechanics, which lead back to the so-called  Ito-Stratonovich dilemma~\cite{brenner1978,landsberg1984,vankampen1988,sattin2008,andreucci2019,bringuier2011,schnitzer1993,jayannavar1995,won2019,lancon2002,milligen2005,bringuier2009,lau2007,sokolov2010,farago2014,serov2020,volpe2010,pesce2013,lanoiseleeDiffusionlimitedReactionsDynamic2018,ryskinSimpleProc1997,escandeWhenCan2007,sandev2015}. In particular, these approaches propose different laws that
generalize linear constitutive relations to account for space-dependent properties
of a physical system. In the problem of mass diffusion---the paradigmatic model of
transport phenomena \cite{Rondoni2021}---the controversy revolves around the Fick's law
\begin{equation}\label{eq:Fick}
  \bm{J}_D = -\mathsf{D} \cdot \nabla\rho,
\end{equation}
which expresses a component of the matter flow $\bm{J}_\mathsf{D}(\bm{x})$ driven
by the gradient of density $\rho(\bm{x})$ through a space-dependent tensor of diffusion coefficients $\mathsf{D}(\bm{x})$, and the alternative Fokker-Planck law
\begin{equation}\label{eq:FP}
  \bm{\tilde{J}}_D = -\nabla\cdot\left( \mathsf{\rb{D}} \rho \right),
\end{equation}
in which the flux is denoted by tilde in contradistinction from the analogous quantity in Eq.~\eqref{eq:Fick}.

Fick's and Fokker-Planck laws can be motivated by deriving a mass diffusion equation from simplified models of microscopic dynamics~\rb{\cite{brenner1978,landsberg1984,vankampen1988,sattin2008,andreucci2019,jayannavar1995,bringuier2011,schnitzer1993,won2019,chacon2007}}. \rb{In presence of an external field $U(\bm{x})$ s}uitable physical assumptions about a system of interest lead either to the Smoluchowski equation,
\begin{equation}\label{eq:SE}
  \frac{\partial\rho}{\partial t} = -\nabla\cdot\bm{J} = \nabla\cdot\left(
    \mathsf{D}\cdot\frac{\nabla U}{k_B T}\rho + \mathsf{D}\cdot\nabla\rho
  \right),
\end{equation}
in which the last term on the right-hand side entails the Fick's law, or to the Fokker-Planck
equation
\begin{equation}\label{eq:FPE}
  \frac{\partial\rho}{\partial t} = -\nabla\cdot\bm{\tilde{J}} = \nabla\cdot\left[
    \mathsf{\rb{D}}\cdot\frac{\nabla\rb{U}}{k_B T}\rho + \nabla\cdot\left(\mathsf{\rb{D}}\rho\right)
  \right],
\end{equation}
implying the Fokker-Planck law. In Eqs.~\eqref{eq:SE}--\eqref{eq:DE}, $k_B$ and $T$ are the Boltzmann constant and
temperature, respectively, whereas $\bm{J}(\bm{x})$ and \rb{$\bm{\tilde{J}}(\bm{x})$ are two formulations of} the total mass flux.

The choice between Eqs.~\eqref{eq:SE} and \eqref{eq:FPE} causes the existing debate:
which law---Eq.~\eqref{eq:Fick} or \eqref{eq:FP}---generalizes linear constitutive relations for inhomogeneous transport
phenomena? Whereas some authors favor the Fokker-Planck approach \cite{lancon2002,milligen2005,won2019},
other suggest that the choice depends entirely on microscopic details of a specific
system of interest \rb{and, therefore, various systems could respond differently to the same external potential $U(x)$}~\cite{landsberg1984,vankampen1988,sattin2008,andreucci2019,sokolov2010}.

\rb{In contrast to the existing approaches, we argue in the present paper that the Smoluchowski and Fokker-Planck pictures are equivalent if one recognizes a distinct physical quantity $\tilde{U}(x)$, \textit{further called the Fokker-Planck potential}, which must be used in Eq.~\eqref{eq:FPE} \textit{in place of} $U(x)$:
\begin{equation}\label{eq:DE}
  \frac{\partial\rho}{\partial t} = -\nabla\cdot\bm{J} = \nabla\cdot\left[
    \mathsf{D}\cdot\frac{\nabla\tilde{U}}{k_B T}\rho + \nabla\cdot\left(\mathsf{D}\rho\right)
  \right],
\end{equation}
Hence one and the same system can be described using either of the two models~\eqref{eq:SE} or \eqref{eq:DE}, whose potential terms are related by
\begin{equation}\label{eq:rules}
  U(\bm{x}) = \tilde{U}(\bm{x}) + k_B T h(\bm{x}),
\end{equation}
in which the (dimensionless) function $h(\bm{x})$ depends on the components of the tensor $\mathsf{D}(\bm{x})$.}

However only the energy function $U(\bm{x})$ can be interpreted as a thermodynamic potential
consistently with the classical statistical mechanics and the Fick's law. The density
$\rho(\bm{x})$ of an equilibrium system must obey the Maxwell-Boltzmann distribution---a reason
why the mass diffusion may be considered as a paradigmatic example of the transport
phenomena \cite{Rondoni2021}. This result is reproduced only for the potential $U(\bm{x})$ in the Smoluchowski equation~\cite{lau2007} and extends to nonequilibrium systems through Onsager's theory \cite{onsager1931I,onsager1931II,onsager1953}.

The Smoluchowski and Fokker-Planck equations can also describe an ensemble of Brownian
particles  exploring an energy landscape with a space-dependent diffusion coefficient~\cite{lau2007}. In such a model the energy function $\tilde{U}(\bm{x})$,
determines the preferred direction of the \rb{microscopic particle} flow. This interpretation is also supported by our analysis of a simplified microscopic model of diffusion (Sec.~\ref{sec:mic}).

In a recent work \cite{belousov2020} we combined stochastic theory of diffusion with molecular-dynamics simulations to investigate mobility of water molecules near the amino-acid surface of a glutamine crystal, which has been implicated in numerous neurodegenerative diseases. In that study, assuming Eq.~\eqref{eq:SE}, the gradient of external potential and the mean first-passage time of the molecules escaping a narrow spatial region were measured, and then used to infer the transverse inhomogeneous diffusion coefficient $D(x)$ as a function of distance $x$ to the crystal's surface.

Here, using the \textit{equivalence} of the Smoluchowski and Fokker-Planck pictures we
propose a new inference method for both, the transverse  potential term, Eq.~\eqref{eq:inferU}, and the space-dependent diffusivity, Eq.~\eqref{eq:inferD}.  
Notably, our method relies only on the first-passage statistics extracted from stochastic trajectories of the system of interest. We thoroughly test  this estimation technique in a model of the B\"uttiker-Landauer ratchet \cite{buttikerTransportConsequenceStatedependent1987,landauerMotionOutNoisy1988}~(Sec.~\ref{sub:ld}).

As an application of the new inference method, we present a study of liquid-water molecules'
diffusion in two systems involving phase boundaries.
We perform molecular-dynamics simulations of water-vapor and water-surfactant interfaces in a slab geometry~\cite{liu2004,liu2005,falciani2020}
(Sec.~\ref{sec:app} and Appendix~\ref{app:md}). Surfactants are essentially soap-like molecules that have an amphiphilic character and form soap films when put in liquid water. Water dynamics in these systems has been shown experimentally to be quite complex~\cite{baryiamesBurstingBubbleMolecular2021}. From our simulations we obtain trajectories of the water molecules and measure their transverse diffusion coefficient as a function of proximity to the phase boundary. Interestingly, the molecule's mobility at the water-vapor interface is enhanced, whereas over a region extending approximately to \SI{3}{nm} from the water-surfactant interface the diffusion slows down by a factor of two relative to the bulk.  The inferred space-dependent diffusivity is compared with  the value obtained using the approach of  Ref.~\cite{belousov2020}  which requires additional measurements of the local particle density. Both methods agree with the diffusion coefficients inferred from the linear regression of local mean-squared displacements (Appendix~\ref{app:msd}).

The extent to which interfaces affect the structure and dynamics of water molecules has been the subject of numerous experimental and theoretical studies~\cite{Fenimore2002,Levy2006,Sharma2018,Laag2017,shadrack2020computational}. While it has been appreciated that biological materials, such as proteins and DNA, typically decrease diffusion of water, the spatial range of this effect has been a subject of numerous debates~\cite{PhilipBall2008}. Over the last decade several molecular-dynamics simulations have shown that near biological interfaces water molecules slow down by a factor of four to seven compared to the bulk liquid~\cite{Laag2017}. Since these interfaces are often characterized by highly heterogeneous chemical environments~\cite{knapp1993}, the notion of a uniform diffusion constant becomes questionable and therefore provides a fertile ground to test and explore our theory.

\section{\label{sec:thr}Theory}
For simplicity we first consider one-dimensional mass transport in equilibrium systems. We return to the general case in Sec.~\ref{sec:mac}, see also Appendices~\ref{app:FP2D}--\ref{app:DT}. As noted in Ref.~\cite[Sec. V]{lau2007}, one-dimensional systems are always
``conservative,'' because Eq.~\eqref{eq:SE} can be converted into \eqref{eq:DE} by
the substitution
\begin{eqnarray}\label{eq:rule}
  U (x)&=& \tilde{U}(x) + k_B T \log [D(x) / D_0],
\end{eqnarray}
in which $D(x)$ is a one-dimensional diffusion coefficient and $D_0$ is an arbitrary constant. Identifying $h(x) = \log [D(x)/D_0]$ we recognize in the above rule a special case of Eq.~\eqref{eq:rules} \rb{which is known in the literature with $\tilde{U}(x)$ referred to as a convection term~\cite{bringuier2011,escandeWhenCan2007}. Such a simple form of $h(\bm{x})$ is however not available for general diffusion tensor in three dimensions (Appendix~\ref{app:DT})}.

Because in equilibrium the total flux must vanish identically, $J(x) \equiv 0$, for Eqs.~\eqref{eq:SE} and \eqref{eq:DE} we have, respectively,
\begin{align}
  &J = -\left(
    \frac{D}{k_B T} \frac{\partial U}{\partial x} \rho
    + D \frac{\partial\rho}{\partial x}
  \right) = 0,\\
  &J = -\left[
    \frac{D}{k_B T} \frac{\partial\tilde{U}}{\partial x} \rho
    + \frac{\partial\left(D\rho\right)}{\partial x}
  \right] = 0,
\end{align}
which in view of Eq.~\eqref{eq:rule} are solved simultaneously by
\begin{equation}\label{eq:1D}
  \rho(x) \propto \me^{-U(x)/k_B T}
  = \frac{1}{D(x)} \me^{-\tilde{U}(x)/k_B T}.
\end{equation}
The equilibrium density $\rho(x)$ is thus given by a Boltzmann factor with the effective potential $U(x)$, as dictated by  classical statistical mechanics.
The same solution expressed in the Fokker-Planck form depends on two space-dependent factors through the functions $D(x)$ and $\tilde{U}(x)$, where the latter cannot be interpreted as the effective potential of an equilibrium system. In Sec.~\ref{sec:mac} we extend this argument to the nonequilibrium case.

Now we proceed with an independent analysis of a simplified microscopic dynamics, which encompasses a physical interpretation of the Fokker-Planck potential. Our analysis also connects the theory of inhomogeneous mass transport with the stochastic thermodynamics of discrete Markov processes. Returning to the macroscopic description
of the problem in Sec.~\ref{sec:mac}, we use a consistency with the classical theory
of equilibrium systems as the \textit{physical} principle that entails Eq.~\eqref{eq:rules}
and imposes additional constraints on the components of a diffusion tensor in two
and more dimensions (Appendix~\ref{app:DT}).

\subsection{\label{sec:mic}Microscopic description}
Consider a simplified model of diffusing particles, whose states are described by a single coordinate $x$. Particles jump from a position $x$ to $x+dx$ with a space-dependent rate $k_+(x)$ and from $x$ to $x-dx$ with a space-dependent rate $k_-(x)$. In a small interval of time $dt$, we may thus define the probabilities of moving to the right $P_+(x)$ or to the left $P_-(x)$:
\begin{eqnarray}
\label{eq:Ps2}
  P_\pm(x) &=&  k_\pm(x) dt,
\end{eqnarray}
 and a \textit{survival}
probability
\begin{equation}\label{eq:Ps}
 P_0(x) =  1 - \Gamma(x) dt,
\end{equation}
in which $\Gamma(x) = k_+(x) + k_-(x)$ is the escape rate \cite{maes2020}.

\rb{In addition we} introduce a \textit{directing function} $\mathcal{L}(x)$, so that the conditional probability of the forward or backward moves, given that a particle in the state $x$
makes a jump in a time interval $dt$, are expressed by
\begin{equation}\label{eq:moves}
\frac{P_\pm(x)}{1 - P_0(x)} = \frac{k_\pm(x)}{\Gamma(x)}
= \frac{\me^{\mathcal{L}(x \pm dx)}}{Z(x)}
\end{equation}
with a normalization factor
$$Z(x) = \me^{\mathcal{L}(x - dx)} + \me^{\mathcal{L}(x + dx)}.$$

\rb{Diffusion equations are routinely derived for the random-walk problems of the kind defined above by starting from a balance equation
for the particles' density\cite{chandrasekhar1943,keller2004,bringuier2011,belousov2016,andreucci2019}:}
\begin{multline}
\rho(x,t+dt) = P_0(x) \rho(x,t) + P_+(x - dx) \rho(x-dx,t)
\\+ P_-(x + dx) \rho(x+dx,t).
\end{multline}
Expanding the left-hand side in power series up to the terms of order $\sim dt$ and
the right-hand side up to $\sim dx^2$ we get
\begin{eqnarray}\label{eq:occupation}
\partial_t \rho &\simeq& \partial_x \left\{
- \frac{dx}{dt} (P_+ - P_-) \rho
+ \partial_x \left[\frac{(P_+ + P_-) dx^2}{2 dt} \rho\right]
\right\} \nonumber\\
&\simeq& \partial_x \left\{
- \Gamma dx^2 \rho \partial_x \mathcal{L} 
+ \partial_x \left[\frac{\Gamma dx^2}{2} \rho\right]
\right\},
\end{eqnarray}
\rb{in which we used}
\begin{eqnarray}
P_{\pm}(x) &=& \Gamma(x)\,\frac{
    \me^{\mathcal{L}(x\pm dx)}
}{
    \me^{\mathcal{L}(x - dx)} + \me^{\mathcal{L}(x + dx)}
}
\nonumber\\
&=& \frac{\Gamma(x)}{2} \left[
    1 \pm dx\,\partial_x \mathcal{L}(x) + O(dx^2)
\right].
\end{eqnarray}

Equation~\eqref{eq:occupation} can be interpreted as a Fokker-Planck diffusion law~\eqref{eq:DE}
$$
\partial_t \rho = \partial_x \left[
D \frac{\partial_x \tilde{U}}{k_B T}\rho
+ \partial_x \left(D\rho\right)
\right],
$$
by identifying the Fokker-Planck potential and the diffusion coefficient  as follows
\begin{eqnarray}\label{eq:Udef}
\tilde{U}(x) &=& - 2 k_B T \mathcal{L}(x),
\\\label{eq:Ddef}
D(x) &=& \Gamma(x) \frac{dx^2}{2}.
\end{eqnarray}

\rb{Now we must show that the Fokker-Planck potential defined by Eq.~\eqref{eq:Udef} is related to the thermodynamic potential $U(x)$ by \eqref{eq:rule}. To do so we apply a decomposition of
the stochastic kinetics Eqs.~\eqref{eq:Ps2}--\eqref{eq:moves} suggested by Refs.~\cite{maes2020,basu2015nonequilibrium,roldan2019exact,baiesi2009nonequilibrium}}. In this formalism, the transition rates read
\begin{equation}\label{eq:frenesy}
k_{\pm} (x) = a(x,x\pm dx) \me^{s(x,x\pm dx)/2}.
\end{equation}
Here $a(x,x\pm dx)$ and $s(x,x\pm dx)$  denote respectively the activity and environmental entropy flow (in $k_{\rm B}$ units) associated with the transition $x\to x\pm dx$. Importantly, the activity is a symmetric property with respect to the exchange of the transition states $a(x,x\pm dx)=a(x\pm dx,x)$ whereas the entropy flow is asymmetric $s(x,x\pm dx)=-s(x\pm dx,x)$
.

Operating with infinitesimal quantities, such as $dt$ and $dx$, in the following
we frequently invoke the \textit{adequality}  $\me^{dq} \simeq 1+dq$
for a small $dq$. For instance, it will be convenient to introduce the survival rate
\begin{eqnarray}\label{eq:rest}
  Q(x) &=& P_0(x) / dt,
\end{eqnarray}
related to
\begin{eqnarray}\label{eq:rest2}
  \Gamma(x) dt &=& 1 - P_0(x) = 1 - Q(x) dt \simeq \me^{-Q(x) dt},
\end{eqnarray}
where we have used the fact that $Q(x)dt$ is a small quantity. In terms of the survival rate and the directing function
the transition rates read\vspace{1ex}
\begin{eqnarray}\label{eq:ks}
k_\pm(x) &\simeq& \frac{1}{2 dt} \me^{-Q(x) dt \pm \partial_x \mathcal{L}(x) dx},\\\label{eq:ks2}
k_\mp(x \pm dx) &\simeq& \frac{1}{2 dt} \me^{-Q(x) dt \mp \partial_x Q(x) dt dx \mp \partial_x \mathcal{L}(x) dx}.
\end{eqnarray}

Using Eqs.~\eqref{eq:Ps2}--\eqref{eq:ks2} we derive the symmetric coefficients of activity and the asymmetric entropy differences
defined as~\cite{maes2020}:
\begin{eqnarray}
a(x,x+dx) &=& \sqrt{k_+(x) k_-(x + dx)}
\nonumber\\
&\simeq& \frac{1}{2 dt} \me^{-Q(x)dt - \frac{\partial_x Q(x) dt dx}{2}}
,\\
s(x,x+dx) &=&
= k_B \ln\frac{k_+(x)}{k_-(x+dx)} \nonumber\\
&\simeq& k_B \partial_x [Q(x) dt + 2 \mathcal{L}(x)] dx \nonumber\\
&=& \partial_x S(x) dx = dS(x),
\end{eqnarray}
in which we recognize the Boltzmann entropy
\begin{equation}\label{eq:Sdef}
S(x) = k_B [ 2 \mathcal{L}(x) + Q(x) dt ]
\end{equation} of a state $x$, as explained shortly.

The equilibrium density of particles must be given by the Boltzmann ansatz \cite{onsager1953}
\begin{equation}\label{eq:Seq}
\rho(x) \propto \me^{S(x)/k_B} \propto \me^{-\frac{U(x)}{k_B T}},
\end{equation}
which satisfies the detailed-balance condition
\begin{equation}\label{eq:balance}
\rho(x) P_+(x) = \rho(x+dx) P_-(x+dx).
\end{equation}
Indeed, on the left-hand side of Eq.~\eqref{eq:balance} we find
\begin{multline}\rho(x) P_+(x) \propto
\me^{-\frac{U(x)}{k_B T}} k_+(x) dt \\\simeq
\frac{1}{2}\me^{\frac{S(x)}{k_B}-Q(x)dt + \partial_x \mathcal{L}(x) dx}
= \frac{1}{2}\me^{2 \mathcal{L}(x) + \partial_x \mathcal{L}(x) dx},
\end{multline}
which equals the right-hand side
\begin{multline}\rho(x+dx) P_-(x + dx) \propto
\me^{-\frac{U(x) + \partial_x U(x) dx}{k_B T}} k_-(x) dt \\=
\frac{1}{2}\me^{\frac{S(x) + \partial_x S(x) dx}{k_B}-Q(x)dt - \partial_x Q(x) dt dx - \partial_x \mathcal{L}(x) dx}
\\\simeq \frac{1}{2}\me^{2 \mathcal{L}(x) + \partial_x \mathcal{L}(x) dx},
\end{multline}
by virtue of Eqs.~\eqref{eq:ks}--\eqref{eq:ks2} and \eqref{eq:Sdef}--\eqref{eq:Seq}.

\rb{Now Eq.~\eqref{eq:rule} follows from Eqs.~\eqref{eq:Udef}--\eqref{eq:Ddef}, \eqref{eq:rest2}, and \eqref{eq:Sdef}--\eqref{eq:Seq}, as had to be demonstrated:
\begin{equation}\label{eq:SUD}
    U(x) = -T S(x) = \tilde{U}(x) - k_B T \ln[D(x)/D_0],
\end{equation}
with $D_0 = dx^2 / (2 dt)$. An extension of the above formalism to nonequilibrium systems (without detailed balance) is also possible (see Appendix~\ref{app:extra}).}

\rb{The above analysis also encompasses a physical interpretation for the Fokker-Planck potential $\tilde{U}(x)$, which is related to the directing function $\mathcal{L}(x)$ through Eq.~\eqref{eq:Udef} and determines the \textit{preferred direction} of the particles' motion. As we show in Sec.~\ref{sec:app}, this interpretation applies also at the mesoscopic scale to the first-passage statistics of the molecules.}

\rb{Finally we remark that the derivation of the Fokker-Planck diffusion equation stated in the beginning of this section can be reproduced for
two and more dimensions, \textit{cf.} Ref.~\cite{bringuier2011}.} For instance, we may
assume that particles move along the coordinates $x_1$ and $x_2$ independently
with rates $\Gamma_1(x_1,x_2)$ and $\Gamma_2(x_2,x_2)$ respectively, whereas the direction of moves is determined by the directing function $\mathcal{L}(x_1,x_2)$ (Appendix~\ref{app:FP2D}).
The independent rates $\Gamma_1$ and $\Gamma_2$ along the two directions represent
a microscopical equivalent of a local reference frame associated with the eigenvectors
of the diffusion tensor (Appendix~\ref{app:DT}), whose diagonal components are then
given by
\begin{equation}\label{eq:Ds}
D_1(x) = \Gamma_1(x) dx^2/2,\quad
D_2(x) = \Gamma_2(x) dx^2/2.
\end{equation}
Because the diffusion tensor is always given by a symmetric positive-definite matrix
in Cartesian coordinates, the diagonal representation must always exist as well as
the eigensystem reference frame.

\subsection{\label{sec:mes} Mesoscopic description}
The theory of Langevin dynamics, which describes inhomogeneous diffusion at the mesoscopic scales and generates the equilibrium density~\eqref{eq:1D}, has been extensively studied by \citet{lau2007}. Ref.~\cite{lau2007} provides a generic stochastic differential equation, which is consistent with the Smoluchowski diffusion~\eqref{eq:SE}:
\begin{equation}\label{eq:LL}
dx= -\left[
\frac{D(x)}{k_B T}\partial_x U(x) - a \partial_x D(x)
\right] dt+ \sqrt{2 D(x)}  dB,
\end{equation}
in which $dB$ is the increment of the Wiener process, i.e. $dB$ is a Gaussian random variable with the zero mean $\langle dB(t) \rangle =0$ and the variance $\langle dB(t)^2\rangle=dt$.  Equation~\eqref{eq:LL} is  interpreted according to the $a$-convention with the parameter $a \in [0,1]$. More precisely  for any real function $f(x)$ we interpret the noise term in~\eqref{eq:LL} as
\begin{multline}
f(x) dB = f\left[a x(t)+ (1-a) x(t+dt)\right]
    \\\times\left[B(t+dt)-B(t)\right],
\end{multline} and similarly for the drift term proportional to $dt$. The choice of the parameter $a$ constitutes the so-called Ito-Stratonovich dilemma. Three conventions commonly used in the literature are $a=1$ (Ito), $a=1/2$ (Fisk-Stratonovich), and $a=0$ (H\"{a}nggi-Klimontovich)~\cite{sokolov2010}. The term $- a \partial_x D(x)$ in Eq.~\eqref{eq:LL} is a noise-induced drift which ensures that for every value of $a$ the stationary state is given by the Boltzmann distribution~\cite{lau2007}. 
Using Eq.~\eqref{eq:rule} we can substitute
$$\partial_x D(x) = D(x) \partial_x \log \frac{D(x)}{D_0}
    = \frac{D(x)}{k_B T} \partial_x\left[U(x)-\tilde{U}(x)\right],$$
into \eqref{eq:LL} to obtain
\begin{multline}\label{eq:LD}
dx = -\frac{D(x)}{k_B T}\partial_x \left[
    (1-a) U(x)  + a \tilde{U}(x)
\right] dt \\+ \sqrt{2 D(x)}  dB.
\end{multline}
Equation~\eqref{eq:LD} explicitly shows that the Ito-Stratonovich dilemma regards merely the correct choice of the potential function. In particular, the Fokker-Planck
potential $\tilde{U}(x)$ should be used with the Ito-Langevin dynamics, whereas the H\"{a}nggi-Klimontovich convention relies on the effective potential $U(x)$. The Fisk-Stratonovich convention mixes the two pictures.

In the following section we will also show that the notion of a Fokker-Planck potential extends to nonequilibrium systems. Therefore the equivalence of the Smoluchowski and Fokker-Planck equations remains valid out of equilibrium,
as well as the resolution of the Ito-Stratonovich dilemma through Eq.~\eqref{eq:LD}.

\subsection{\label{sec:mac}Macroscopic description}

To generalize Eq.~\eqref{eq:rule} to Eq.~\eqref{eq:rules} for multidimensional systems
we consider first the equilibrium case. In the Einstein summation notation components
of a $\nu$-dimensional mass flux assume then the following Fokker-Planck form
\begin{equation}\label{eq:Ji}
  J_i = -\left[
    \mathsf{D}_{ij} \frac{\partial_j \tilde{U}}{k_B T}\rho + \partial_j (\mathsf{D}_{ij} \rho)
  \right] = 0.
\end{equation}
The equilibrium solution
\begin{equation}\label{eq:eU}
  \rho(\bm{x}) \propto \me^{-\frac{U(\bm{x})}{k_B T}}
\end{equation}
substituted into Eq.~\eqref{eq:Ji} yields a set of $\nu$ equations for $j=1,2,...,\nu$
that must hold simultaneously:
\begin{equation}\label{eq:Q}
  k_B T \partial_i h = \partial_i (U - \tilde{U})
    = k_B T (\mathsf{D}^{-1})_{ij} \partial_k \mathsf{D}_{jk}
\end{equation}
in which we have identified the scalar function $h(\bm{x}) = [U(\bm{x})-\tilde{U}(\bm{x}] / (k_B T)$.

From Eq.~\eqref{eq:Q}, which entails Eq.~\eqref{eq:rules}, we deduce that for the equilibrium solution Eq.~\eqref{eq:eU} to exist the components of the diffusion tensor $\mathsf{D}(\bm{x})$ must satisfy certain constraints. In particular, the second derivatives of $h(\bm{x})$ require the following to hold for $i \ne j$:
\begin{eqnarray}\label{eq:ddQ}
  \partial_i [(\mathsf{D}^{-1})_{jk} \partial_l \mathsf{D}_{kl}] &=& \partial_i\partial_j h \nonumber\\
  &=& \partial_j\partial_i h
  = \partial_j [(\mathsf{D}^{-1})_{ik} \partial_l \mathsf{D}_{kl}].
\end{eqnarray}
Although we cannot provide a general form of $h(\bm{x})$, in Appendix~\ref{app:DT}
we derive it for a diagonal diffusion tensor and discuss the consequences of Eq.~\eqref{eq:Q}
and \eqref{eq:ddQ} in two and three dimensions.

Out of equilibrium we may not rely on the Maxwell-Boltzmann statistics. We therefore
should adopt the Onsager's approach, in which the Boltzmann entropy $S(\bm{x})$ supplants
the potential $U(\bm{x})$ \cite{onsager1931I,onsager1931II,onsager1953}:
\begin{equation}\label{eq:eB}
  \rho(\bm{x}) \propto \me^{S(\bm{x})/k_B},
\end{equation}
\textit{cf.} \eqref{eq:Seq}. With general nonequilibrium forces $\bm{F}$ Eq.~\eqref{eq:DE} assumes the form
\begin{equation}\label{eq:gen}
  \partial_t \rho = -\nabla\cdot \bm{J} = -\nabla\cdot\left[
    \bm{F}\rho - \nabla\cdot\left(\mathsf{D} \rho\right)
  \right].
\end{equation}
in which components of the flux $\bm{J}$ do not vanish. Following the approach of Refs.~\cite{sanmiguel1980,schimansky1985,ebeling1986} we express the flux using a skew-symmetric matrix $\mathsf{d}$ with elements satisfying
$\mathsf{d}_{ij} = -\mathsf{d}_{ji}$:
\begin{equation}\label{eq:equivalence}
  \bm{J} = \nabla\cdot\left(\mathsf{d} \rho\right)
    = \bm{F}\rho - \nabla\cdot\left(\mathsf{D} \rho\right).
\end{equation}
By substituting Eq.~\eqref{eq:eB} into \eqref{eq:equivalence} we find
\begin{equation}\label{eq:flow}
  \bm{F} - \nabla\cdot\mathsf{D} - \mathsf{D}\cdot\frac{\nabla S}{k_B}
  - \nabla\cdot\mathsf{d} - \mathsf{d}\cdot\frac{\nabla S}{k_B} = 0.
\end{equation}

Consider a vector of parameters $\bm{\epsilon}$ that characterize the system's departure
from equilibrium. In a linear nonequilibrium regime we should observe
\begin{align}\label{eq:expS}
  S =& -\frac{U}{T} + \bm{\epsilon} \cdot \partial_{\bm{\epsilon}} S  + O(|\epsilon|^2),
  \\\label{eq:Fe}
  \bm{F} =& -\mathsf{D}\cdot\frac{\nabla\tilde{U}}{k_B T}
    + \bm{\epsilon} \cdot \partial_{\bm{\epsilon}} \bm{F}
    + O(|\epsilon|^2),
  \\\label{eq:expd}
  \mathsf{d} =& \bm{\epsilon}\cdot\partial_{\bm{\epsilon}}\mathsf{d} + O(|\epsilon|^2),
\end{align}
in which we use the fact that $\mathsf{d}$ vanishes in equilibrium $|\epsilon| = 0$.
The tensor $\partial_{\bm{\epsilon}} \bm{F}$ represents here the linear Onsager coefficients for the components of the matter flow conjugate to the parameter $\bm{\epsilon}$. Substitution
of Eqs.~\eqref{eq:expS}--\eqref{eq:expd} into \eqref{eq:flow} yields
\begin{multline}\label{eq:perturb}
  -\left(
    \mathsf{D}\cdot\frac{\nabla\tilde{U}}{k_B T} + \nabla\cdot\mathsf{D}
    - \mathsf{D}\cdot\frac{\nabla U}{k_B T}
  \right)\\
  + \bm{\epsilon}\cdot\partial_{\bm{\epsilon}} \left(
    \bm{F} - \mathsf{D}\cdot\frac{\nabla S}{k_B}
    - \nabla\cdot\mathsf{d} + \mathsf{d}\cdot\frac{\nabla U}{k_B T}
  \right) = 0,
\end{multline}
with $\bm{\epsilon}$ regarded as a small perturbation of an equilibrium state. On the left-hand side of Eq.~\eqref{eq:perturb} the first of the two terms, which
must both vanish identically, ensures the consistency with the equilibrium theory
\begin{equation} \nabla U = \nabla\tilde{U} + k_B T \, \mathsf{D}^{-1}\cdot\nabla\cdot\mathsf{D}.   \end{equation}
From this condition we derive Eq.~\eqref{eq:Q} in the vector form
\begin{equation}
       k_B T \nabla h = \nabla (U - \tilde{U})
    = k_B T \mathsf{D}^{-1}\cdot\nabla\cdot\mathsf{D},
\end{equation}
which confirms our claim in Eq.~\eqref{eq:rules}. We may also identify in Eq.~\eqref{eq:perturb} the Fick's law
\begin{equation} \bm{J}_\mathsf{D} = \mathsf{D}\cdot\left(
  \frac{\nabla U}{k_B T} - \bm{\epsilon}\cdot\partial_{\bm{\epsilon}} \frac{\nabla S}{k_B}
\right) \rho = -\mathsf{D} \cdot \nabla \rho.\end{equation}

\section{\label{sec:app} Inferring space-dependent diffusion from first-passage statistics}

\begin{figure}[ht!]
    \centering
    \includegraphics[width=0.85\columnwidth]{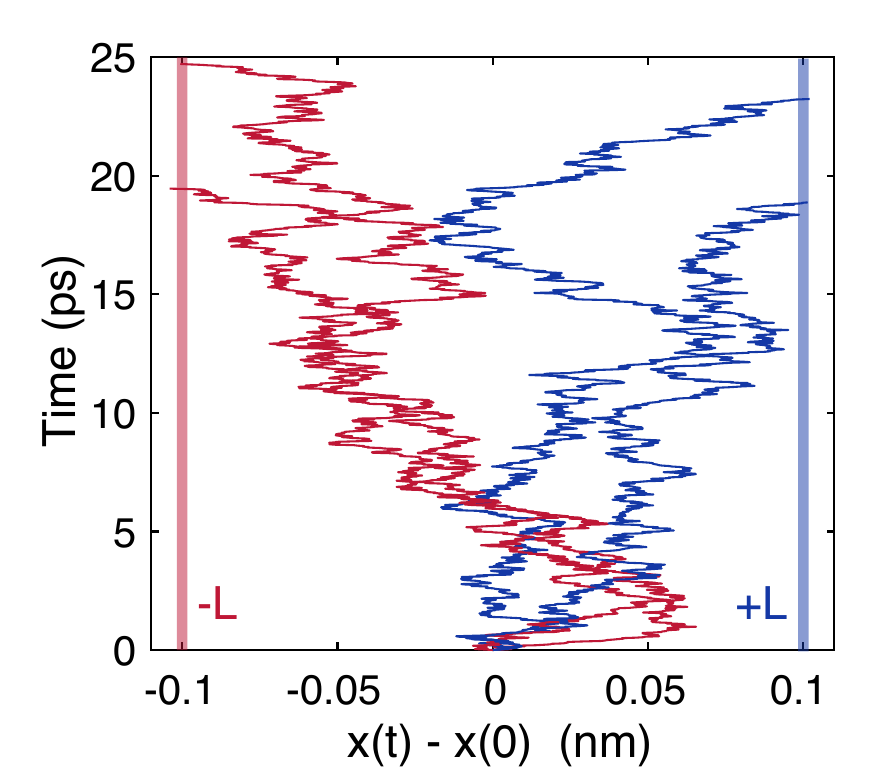}
    \caption{
        Illustration of first passage events in time series of four non-interacting particles. Initially all the four particles are located at $x = \SI{0}{nm}$. The time series of their positions are truncated at the respective first-passage events---when the particles' trajectories cross one of the boundaries at $x-L = \SI{-0.1}{nm}$ or at $x+L = \SI{0.1}{nm}$ ($L=\SI{0.1}{nm}$) for the first time. The red-colored series end with the negative first-passage events at $x-L$, whereas the blue-colored ones---with the positive event at $x+L$. The first-passage time can be read out from the vertical time axis at the final point of the trajectories.
    }
    \label{fig:sketch}
\end{figure}

Molecular-dynamics methods, which allow for estimation of fluids' space-dependent diffusion coefficients in presence of an external potential, have been a topic of several publications \cite{liu2004,hummer2005,sedlmeier2011,olivares2013,belousov2020}.
In particular, some of the latest approaches~\cite{sedlmeier2011,olivares2013,belousov2020}, which offer a relatively simple computational scheme, rely on the information extracted from statistics of first-passage times---a time elapsed until a certain condition, typically a passage to a given spatial region, is satisfied~\cite{redner2001guide,dy2008first,metzler2014first,majumdar2007brownian,forster2021exact}.
By combining the theory presented in Sec.~\ref{sec:thr} with the approach formulated in Ref.~\cite{belousov2020}, we develop a new method for measuring the transverse component of a diffusion tensor.

First we provide definitions relevant for a diffusing particle in one dimension. In what follows, we consider the first-passage time of a molecule, which escapes a symmetric interval of length $2L$ centered at the initial position $x_0=x(0)$. See Fig.~\ref{fig:sketch} for an illustration. Such a first-passage time is formally defined as
\begin{equation}
\tau (x_0) = \inf \{ t\geq 0  \;:\; |x(t)-x_0| \geq L \}.
\end{equation}
Note that $\tau(x_0)$ is a positive random variable which takes a different value for each trajectory $x(t)$. The mean first-passage time is then given by
\begin{equation}
 \langle     \tau (x_0) \rangle   = \int_0^\infty  d\tau\, \tau\, P\left[ \tau (x_0)=\tau\right],
\end{equation}where $P\left[ \tau (x_0)=\tau\right]$ is the probability that the particle exits from the spatial region
of interest $[x_0-L,x_0+L]$ within a time interval $[\tau,\tau+d\tau]$.

In addition the first-passage events defined above can be characterized as positive and negative, when the diffusing molecule escapes, respectively, through the positive or negative end of the interval $[x_0-L,x_0+L]$.
Because in this particular problem particles always exit from the region of interest in a finite time, i.e. $\tau (x_0) < \infty$ for all trajectories, the probabilities of these events $P_+(x_0)<1$ and $P_-(x_0)<1$ obey a relation
\begin{equation}
P_+(x_0) + P_-(x_0)=1.
\end{equation}
\vfill\null

No closed-form expression is known in general for the probabilities and the mean exit time of the above first-passage events in terms of $D(x)$, $U(x)$, and $\tilde{U}(x)$. On the other hand, for first-passage intervals of a small length~$L$, one can approximate the dynamics of a molecule near $x=x_0$ by a diffusion process with a drift proportional to the local gradient $\partial_x U(x_0)$ and a constant mobility $D(x_0)/(k_B T)$. Within this approximation one can relate
the local first-passage statistics to the space-dependent potential and diffusion coefficient by using
the analytical expressions for $\langle\tau(x_0)\rangle$, $P_+(x_0)$, and $P_-(x_0)$ in a drift-diffusion process~\cite{[Chapter 3 in ]goel}.
 
In Ref.~\cite{belousov2020}  it was shown that, assuming a  Smoluchowski diffusion equation, the effective potential $U(x)$ can be reconstructed from the steady-state density $\rho(x)$ of the diffusing particles through $U(x)=-k_{\rm B}T\ln\rho(x)$. The coefficient $D(x)$ can then be inferred from two quantities: (i) the gradient of the inferred effective potential $\partial_x U(x)$; and (ii) the mean exit time $ \langle\tau(x)\rangle$ of particles which escape from a small interval $[x-L,x+L]$.
 
 In the Fokker-Planck picture, however, the numerical differentiation of $U(x)$ and, therefore, the estimation of the local density are unnecessary, as we show shortly. This alternative approach significantly simplifies the computational aspects of the inference procedure. Indeed, the probabilities of the positive and negative first-passage events of an overdamped Brownian particle described by Eq.~\eqref{eq:LD} are determined by the Fokker-Planck potential~\cite{[Chapter 3 in ]goel}:
\begin{eqnarray}\label{eq:fpp}
  P_+(x) &=& \frac{\displaystyle\int_x^{x+L} \frac{dy}{D(y)} \me^{-\frac{U(y)}{k_B T}}}{
        \displaystyle\int_{x-L}^{x+L}  \frac{dy}{D(y)} \me^{-\frac{U(y)}{k_B T}}
    }
    = \frac{\displaystyle\int_x^{x+L} dy \me^{-\frac{\tilde{U}(y)}{k_B T}}}{
    \displaystyle\int_{x-L}^{x+L} dy \me^{-\frac{\tilde{U}(y)}{k_B T}}
  },\quad\\\label{eq:fpm}
  P_-(x) &=& \frac{\displaystyle\int_{x-L}^x
    \frac{dy}{D(y)} \me^{-\frac{U(y)}{k_B T}}}{
    \displaystyle\int_{x-L}^{x+L} \frac{dy}{D(y)} \me^{-\frac{U(y)}{k_B T}}
  }
    = \frac{\displaystyle\int_{x-L}^x dy \me^{-\frac{\tilde{U}(y)}{k_B T}}}{
    \displaystyle\int_{x-L}^{x+L} dy \me^{-\frac{\tilde{U}(y)}{k_B T}}
  },\quad\;
\end{eqnarray}
which hold due to Eq.~\eqref{eq:rule}.
As in Ref.~\cite[Appendix C]{belousov2020} we assume that over a small interval $[x-L,x+L]$ the diffusion coefficient and the slope of the Fokker-Planck potential are approximatively constant. In other words, for $y\in [x-L,x+L]$ we pose
    \begin{eqnarray}\label{eq:Dapx}
        D(y) &\simeq& D(x) + O(y-x),
    \\\label{eq:Uapx}
    \partial_x\tilde{U}(y) &\simeq&
    \partial_x\tilde{U}(x) + O(y-x),
    \end{eqnarray}
\rb{which are accurate for small gradients $|\partial_x\tilde{U}(y)|$ and small $L$.}
Dividing Eq.~\eqref{eq:fpm} by \eqref{eq:fpp} and  using Eq.~\eqref{eq:Uapx}, we get
\begin{equation}\label{eq:inferU}
  \partial_x \tilde{U}(x) \simeq \frac{k_B T}{L} \ln\left[ \frac{P_-(x)}{P_+(x)}\right],
\end{equation}
cf. Eq.~\eqref{eq:moves} and \eqref{eq:Udef}. To find the diffusion coefficient we use the formula for the mean exit time~\cite{[Chapter 3 in ]goel}
\begin{eqnarray}
   \langle \tau(x) \rangle &=& \int_x^{x+L} \frac{dy}{D(y)} \int_{x-L}^y dz\, \me^{-\frac{U(z)-U(y)}{k_B T}}
        \nonumber\\
        &-& P_-(x) \int_{x-L}^{x+L} \frac{dy}{D(y)} \int_{x-L}^y dz\, \me^{-\frac{U(z)-U(y)}{k_B T}}
        \nonumber\\
        &=& \int_x^{x+L} dy \int_{x-L}^y \frac{dz}{D(z)}\, \me^{-\frac{\tilde{U}(z)-\tilde{U}(y)}{k_B T}}
        \nonumber\\
        &-& P_-(x) \int_{x-L}^{x+L} dy \int_{x-L}^y \frac{dz}{D(z)}\, \me^{-\frac{\tilde{U}(z)-\tilde{U}(y)}{k_B T}}
        \nonumber\\
        &\simeq& \frac{k_B T L}{D(x)\partial_x\tilde{U}(x)} \tanh\left[ \frac{L \partial_x\tilde{U}(x)}{2 k_B T} \right],
\end{eqnarray}
in which we again used the approximations~\eqref{eq:Dapx} and \eqref{eq:Uapx}. From the last expression, further simplified through Eq.~\eqref{eq:inferU}, $D(x)$ can be expressed as
\begin{equation}\label{eq:inferD}
  D(x) \simeq \frac{L^2}{\langle\tau(x)\rangle}
     \frac{P_+(x)-P_-(x)}{
        \ln \left[P_+(x)/P_-(x)\right].
    }
\end{equation}Remarkably, Eq.~\eqref{eq:inferD} implies that the local diffusion coefficient can be determined from the statistics of first-passage
events alone.      In the following subsection we verify Eqs.~\eqref{eq:inferU} and \eqref{eq:inferD} for a one-dimensional stochastic model of diffusion. Then in Sec.~\ref{sub:md} we apply our inference method to molecular-dynamics simulations of two soft-matter systems with phase boundaries.

\subsection{\label{sub:ld} Application I: B\"uttiker-Landauer ratchet}
\begin{figure}[t]
    \centering
    \includegraphics[width=\columnwidth]{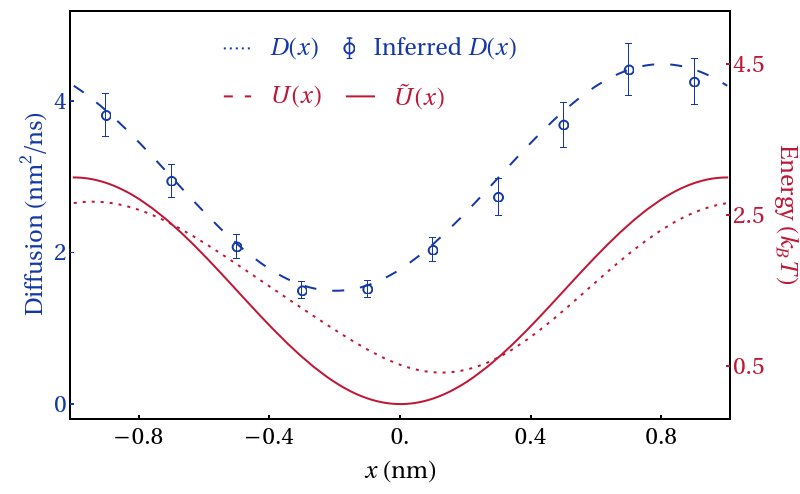}
    \caption{
        Estimation of the diffusion coefficient in a B\"uttiker-like ratchet. Space-dependent diffusion coefficient, plotted together with the Smoluchowski and Fokker-Planck potentials, is compared with the estimates made from our simulations using Eq.~\eqref{eq:inferD}. Error bars are given by three standard deviations.
        Simulation parameter values are: $U_0 = 3 k_B T$, $D_0 = \SI{3}{nm^2/ns}$, $\alpha=0.5$, $\phi=\SI{0.2}{nm}$. Checks for first-passage events with $L=\SI{0.1}{nm}$ were performed at each simulation time step---every $\SI{2}{fs}$. A total of 1000 such events was acquired for each value of the initial position $x$.}
    \label{fig:toy}
\end{figure}

First we test Eqs.~\eqref{eq:inferU} and~\eqref{eq:inferD} in a one-dimensional stochastic model of inhomogeneous transport. In particular, we consider  the so-called  B\"uttiker-Landauer ratchet~\cite{buttikerTransportConsequenceStatedependent1987,landauerMotionOutNoisy1988} with a periodic potential well and diffusion coefficient (Fig.~\ref{fig:toy}):
\begin{eqnarray}
    U(x) &=& \frac{U_0}{2} \left[ 1 - \cos(2 \pi x) \right],\\
    D(x) &=& D_0 \left\{ 1 + \alpha \cos\left[ 2 \pi (x - \phi) \right] \right\},
\end{eqnarray}
in which $U_0 > 0$, $D_0 > 0$, $|\alpha| < 1$, $|\phi| < 1$ are constant parameters. The parameter $U_0$ controls the energy barrier of the potential well, whereas $D_0$ is the diffusion scale. The amplitude of $D(x)$ is modulated by $\alpha$ and its relative phase shift with respect to $U(x)$---by $\phi$. The Ito-Langevin Eq.~\eqref{eq:LD} for this model reads
\begin{equation}\label{eq:eom}
    dx = - \frac{D(x)}{k_B T} \frac{d\tilde{U}(x)}{dx}dt + \sqrt{2 D(x)} dB
\end{equation}
with $\tilde{U}(x) = U(x) - k_B T \ln\left[ D(x) / D_0 \right]$; $B(t)$ denotes  the standard Wiener process.

\begin{figure}[ht!]
    \centering
    \includegraphics[width=\columnwidth]{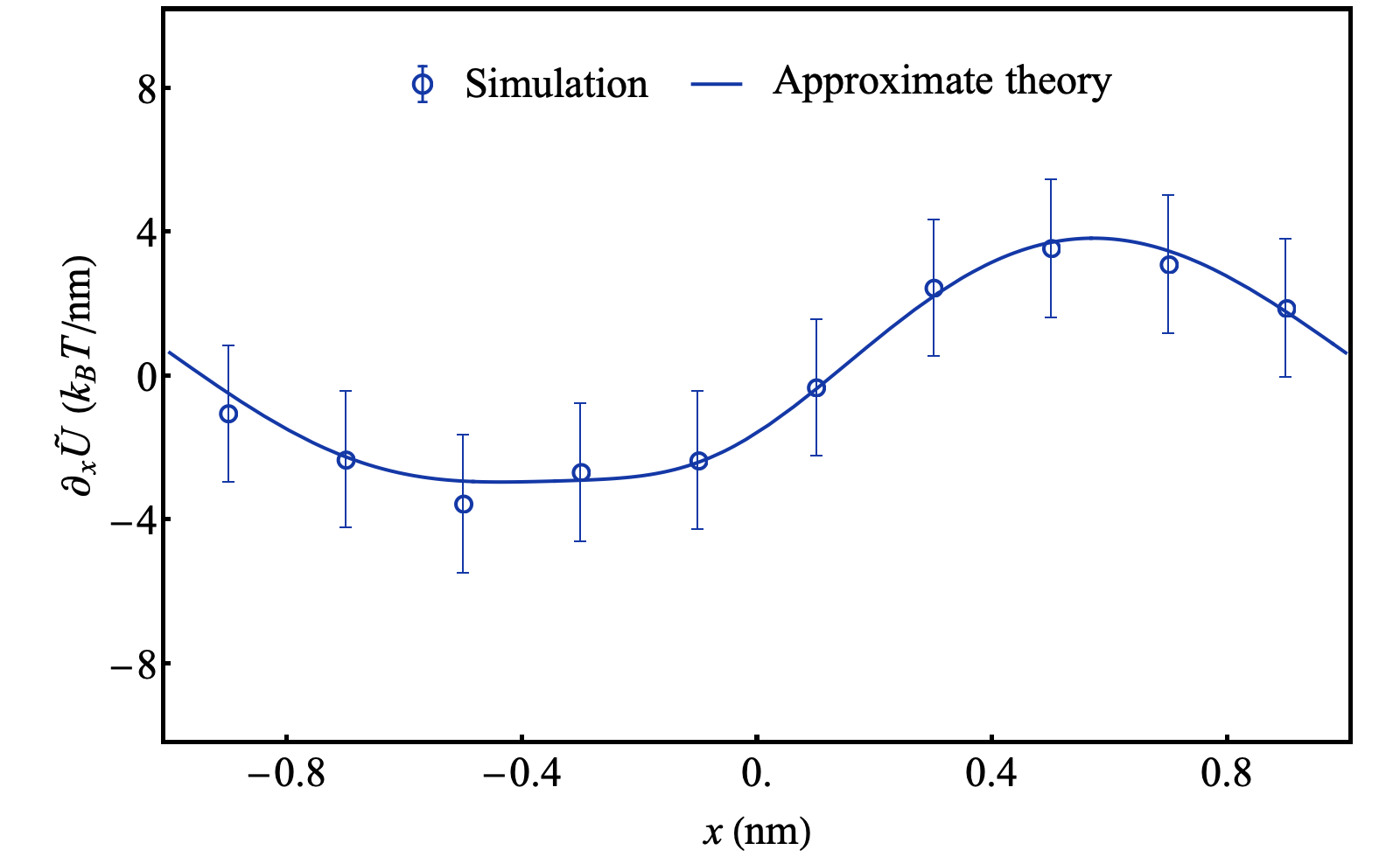}
    \caption{
        Estimation of the gradient of the Fokker-Planck potential in a B\"uttiker-like ratchet. The slope of the Fokker-Planck potential is compared with the approximate Eq.~\eqref{eq:inferU} applied to the simulation data.
        Error bars and simulation parameters are the same as those in Fig.~\ref{fig:toy}.
     }
    \label{fig:log}
\end{figure}

In our example we simulate the system with parameter values, which are similar in magnitude   to those of Sec.~\ref{sub:md}: $U_0 = 3 k_B T$, $D_0 = \SI{3}{nm^2/ns}$, $\alpha=0.5$, $\phi=\SI{0.2}{nm}$, $L=\SI{0.1}{nm}$. The equation of motion \eqref{eq:eom} was integrated using the explicit Euler scheme with a time step $\SI{2}{fs}$. We sampled first-passage events for 1000 initial conditions uniformly distributed in each region of interest $[x-L,x+L]$. Figures~\ref{fig:toy} and \ref{fig:log} show that the first-passage statistics renders reliable estimates for the diffusion coefficient and the gradient of the Fokker-Planck potentials without any fitting parameters: within the statistical uncertainties Eqs.~\eqref{eq:inferU} and~\eqref{eq:inferD} accurately match the simulated profiles of $\partial_x U(x)$ and $D(x)$.

\subsection{\label{sub:md} Application II: Soft-matter interfaces}
In this section we study dynamics of water molecules near two biologically relevant soft-matter interfaces, by using our method based on first-passage statistics. Inspired by recent works~\cite{liu2004,sedlmeier2011,olivares2013,belousov2020}, we model the trajectory of a single molecule by an overdamped Langevin equation in the presence of  external potential and inhomogeneous diffusion coefficient, which
induces multiplicative noise. In one of the systems that we consider, few thousands of water molecules interact with a biological surface through electrostatic and Van-der-Waals forces. The effective Langevin dynamics accounts therefore
for the effective potential and space-dependent diffusivity, which
emerge from microscopic potential-energy interactions, entropic forces and the underlying geometry of the surface under consideration.

\begin{figure}[ht!]
  \includegraphics[width=0.9\columnwidth]{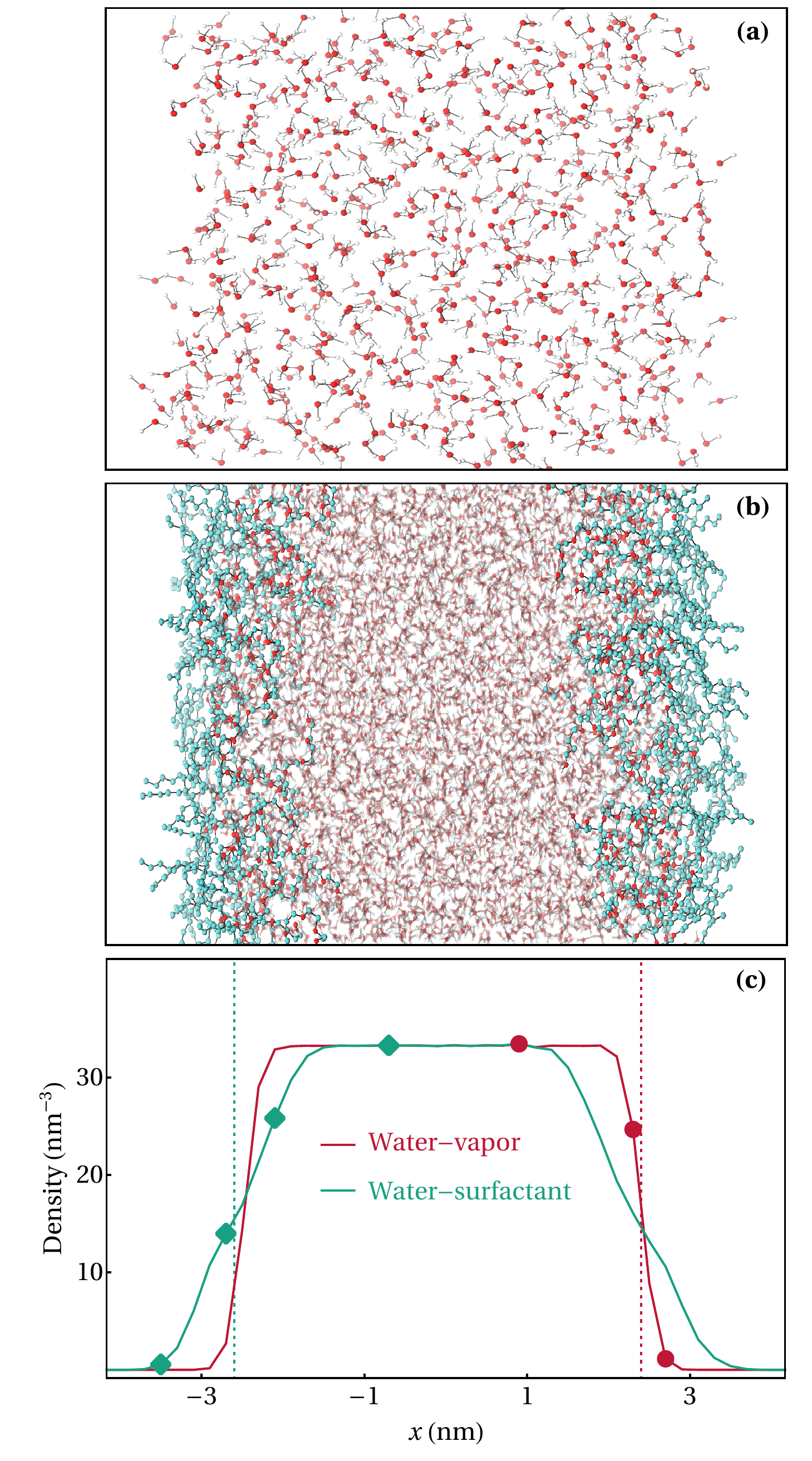}
  \caption{
    \label{fig:md} Water-vapor and water-surfactant interfaces in molecular-dynamics simulations.
    Panel (a): A snapshot of the $\ce{H2O}$ molecules in the water-vapor system, with red-colored oxygen atoms and white-colored hydrogen atoms.
    Panel (b): A snapshot with the surfactant's carbon chains (green-colored) and $\ce{H2O}$ molecules in the water-surfactant system.
    Panel (c): \rb{Equilibrium} number density of water molecules at the center of the simulation box for the water-vapor (red) and water-surfactant (green) interfaces. The red dotted vertical line indicates a position of the Gibbs-dividing interface only in the right part of the simulation
    box for the water-vapor system. Likewise the green dotted vertical line delimits one of the two phase boundaries between water and surfactant in the left part of the simulated system. Red circles and green triangles indicate
    locations inspected in Fig.~\ref{fig:msd} for water-vapor and water-surfactant interfaces respectively. \rb{The equilibrium density of molecules is reconstructed from  histograms with bin size $0.2$\si{nm} extracted from a simulation of total duration $0.5$ns with time step $10$fs.} 
  }
\end{figure}

In particular,  here we consider an effective one-dimensional description of two systems along the
coordinate axis $x$, which is perpendicular to surfaces separating liquid water
from vapor---in the first case [Fig.~\ref{fig:md}(a)]---and from a surfactant layer---in the second case [Fig.~\ref{fig:md}(b)]. Both systems were simulated using a molecular-dynamics package GROMACS in a slab geometry with periodic boundaries (Appendix~\ref{app:md}).
In the {\em water-vapor} simulations a layer of liquid water, which occupies the center of an empty $\SI{300}{nm}\times\SI{5}{nm}\times\SI{5}{nm}$ box, is surrounded from left and right by two gas-liquid interfaces. In the {\em water-surfactant} system of the same size,
two phase boundaries are stabilized by soap films,
which  consist of a surfactant monolayer---molecules of hexaethylene glycol monododecyl ether with a stoichiometric formula $\ce{C6H12}$.

 First we discuss a static property---the local density
 extracted from our molecular-dynamics simulations after equillibration. Compared with the water-surfactant interface, the liquid-vapor contact produces a sharper profile of the water density $\rho(x)$ along the symmetry axis $x$ [Fig.~\ref{fig:md}(c)]. Such space-dependent features of the density profiles
 can be accompanied by a heterogeneous diffusion of water molecules near the phase boundaries. In fact, during a simulation run the molecules may pass across regions, in which their local transport properties vary significantly.

 Now we describe in more details our analysis of the molecular-dynamics data.
The positions of water-molecules' oxygen atoms were acquired from simulated trajectories of a total duration \SI{0.5}{ns}, sampled with a time step $\Delta t= \SI{10}{fs}$. Centers of nonoverlapping regions of interests $[x_i-L,x_i+L]$ of width $2L$ enumerated by an integer~$i$, were distributed over an equidistant grid. When we detect that after $n$ simulation steps a molecule initially located at $x\in[x_i - L, x_i+L]$ has traversed a distance $L=\SI{1}{\textup{\AA}}$, the first-passage time $\tau(x_i) = (n-1)\Delta t +  \Delta t/2 = (n - 1/2) \Delta t$ is recorded.
  The initial coordinates $x$ sampled from the original trajectory were separated by \SI{5}{ps} to reduce statistical correlations between two consecutive first passage events.
  Up to a maximum of about $10^4$ of such events were thus obtained in the most densely occupied regions of interest. Diffusion coefficients were estimated only in the regions of interest with at least $100$ observations---a threshold introduced to ensure sufficient statistics for inference.
  
\begin{figure}[t]
  \includegraphics[width=\columnwidth]{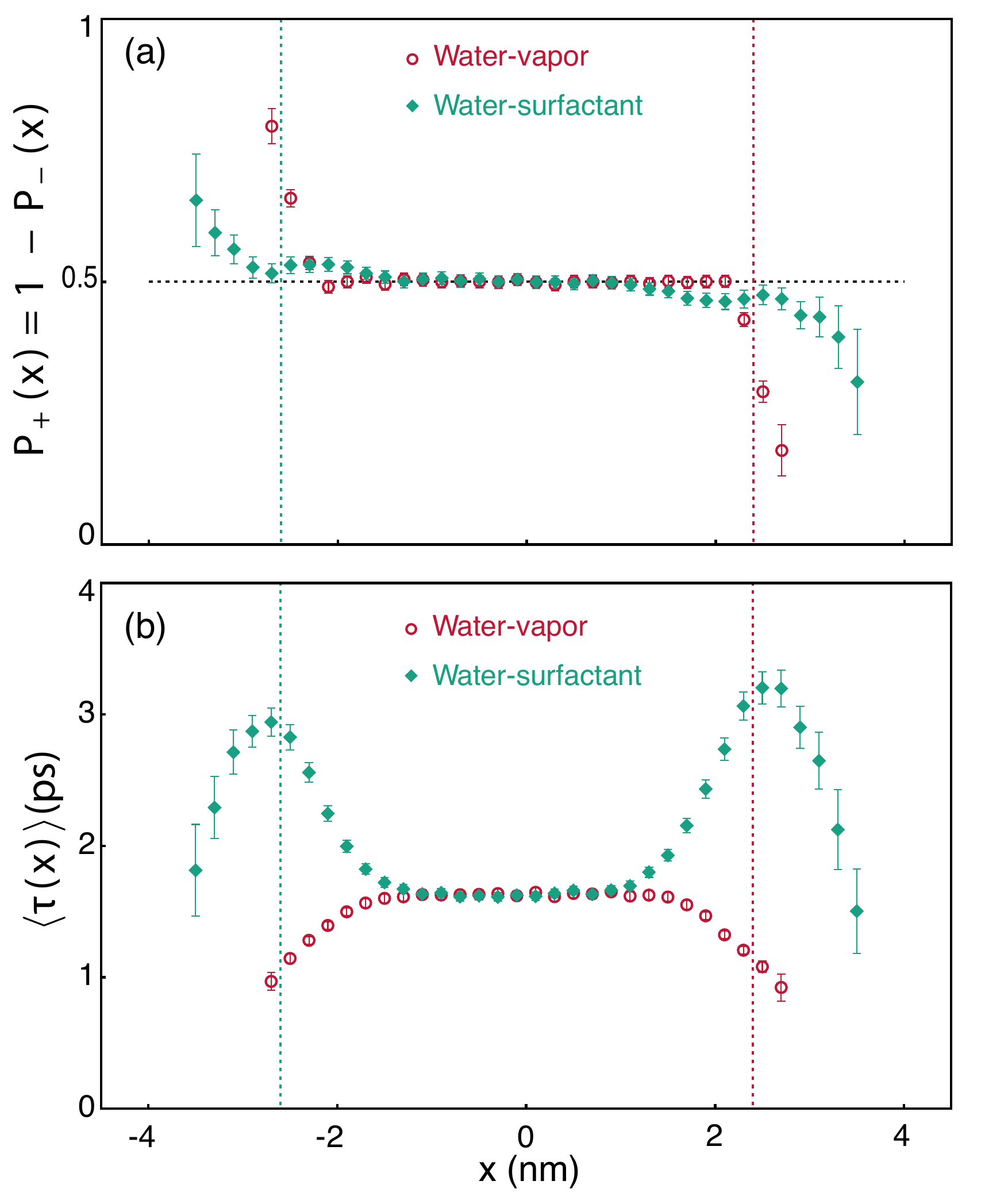}
  \caption{
    \label{fig:fps} First-passage statistics of \ce{H2O} molecules in the water-vapor (red) and water-surfactant (green) systems.
     (a) Conditional probability $P_+(x)$ for water molecules near $x$ to first escape the interval $[x-L,x+L]$  from $x+L$. A first-passage probability  $P_+(x) > 1/2$ ($P_+(x)<1/2$) indicates an effective force acting on the molecules at a given point $x$ in the positive (negative) direction of the axis.
    (b) Mean escape time $\langle \tau(x)\rangle$ from the interval $[x-L,x+L]$ quantifies the molecules' mobility. Larger values of $\langle\tau(x)\rangle$ correspond to a slower diffusive kinetics.
    Error bars are given by three standard deviations. Red and green dotted lines indicate the Gibbs dividing interfaces as described in Fig.~\ref{fig:md}. See Sec.~\ref{sub:md} for further details on the first-passage-time analysis.
  }
\end{figure}

 Figure~\ref{fig:fps} reports the first-passage statistics extracted from our molecular-dynamics data. At distances up to $\sim \SI{1.5}{nm}$ from the Gibbs dividing interfaces the escape probabilities in the direction towards and from the phase boundary are equal in the two simulated systems, $P_+(x)=P_-(x)=1/2$ [Fig.~\ref{fig:fps}(a)]. Closer to the interface, we observe a preferred direction of motion into the bulk liquid, i.e. $P_+(x)>P_-(x)$ ($P_+(x)<P_-(x)$) for $x<0$ ($x>0$). Interestingly, the mean exit time $\langle\tau(x)\rangle $ of water molecules near the water-vapor and water-surfactant phase boundaries present a remarkable qualitative difference [Fig.~\ref{fig:fps}(b)]: $\langle\tau(x)\rangle $, which always decreases with distance from the water-vapor interface, changes nonmonotonously near the water-surfactant contact. In the case of water-vapor simulations the mean exit time $\langle\tau(x)\rangle$, as a function of position, reproduces the shape of the density profile $\rho(x)$ and, therefore, implies that water molecules are faster in less crowded regions as expected. However near the surfactant monolayer no direct correlation between the escape kinetics and the local density of water molecules is observed, cf. Fig.~\ref{fig:md}(c) and Fig.~\ref{fig:fps}(b) which reveal an underlying hetereogeneous chemical environment dicussed below.
 
\begin{figure}[ht!]
\includegraphics[width=0.97\columnwidth]{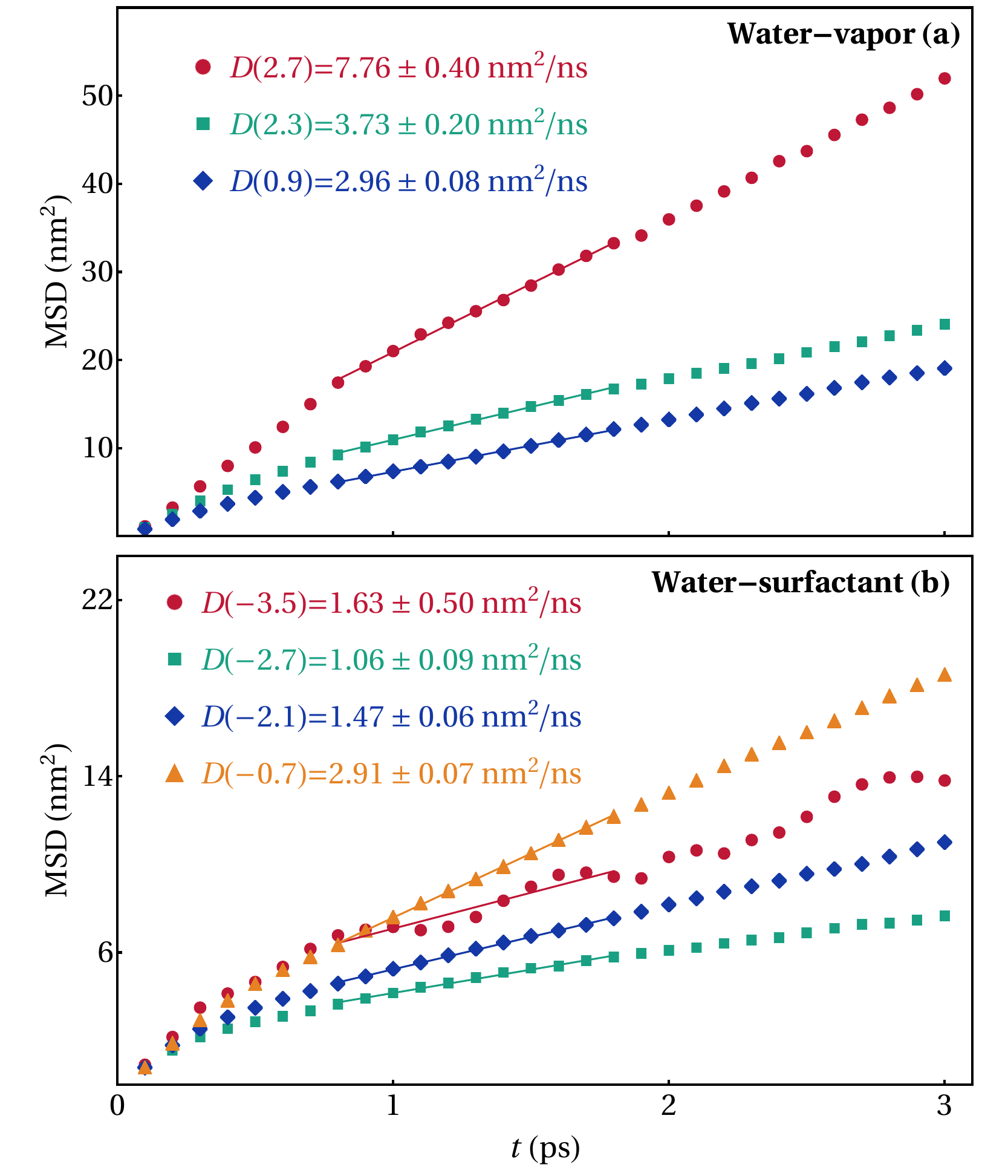}
\caption{
\label{fig:msd} Fitting local mean squared displacements (MSD) of water molecules observed
in our molecular-dynamics simulations for selected locations, which are also indicated
in the density profile in Fig.~\ref{fig:md}(c): (a) water-vapor system and (b) water-surfactant system. The legend shows the inferred values of the diffusion coefficient at different positions $x$ (in \si{nm}), as obtained by a linear regression of the MSDs vs. time within selected intervals (solid lines).
}
\end{figure}

By using three different methods we infer the transverse space-dependent diffusion $D(x)$ of water molecules from our simulation data:
\begin{itemize}
    \item Method I uses Eq.~\eqref{eq:inferD} in terms of escape probabilities and mean exit times.
    \item Method II extracts $D(x)$ from the local density and the first-passage statistics, as prescribed by Eq.~(3) in Ref.~\cite{belousov2020}.
    \item Method III relies on a linear regression of local mean squared displacements of water molecules, which transiently pass through the regions of interest $[x_i-L,x_i+L]$ (see Appendix~\ref{app:msd} for details).
\end{itemize} In Methods I and II we use the results of first-passage analysis already introduced in this section and in Ref.~\cite{belousov2020}. Method III works as follows: using the same partition of the $x$ axis as in the first-passage approach, we collect the molecules' displacements as a function of time and position. On the scales of the molecules' mean first-passage times ($t<\SI{3}{ps}$), we observe a linear trend of mean squared displacements originating from the same region of interest (Fig.~\ref{fig:msd}). By fitting these data we estimate the local diffusion coefficient (Appendix~\ref{app:msd}).

\begin{figure}[ht!]
\centering
  \includegraphics[width=0.95\columnwidth]{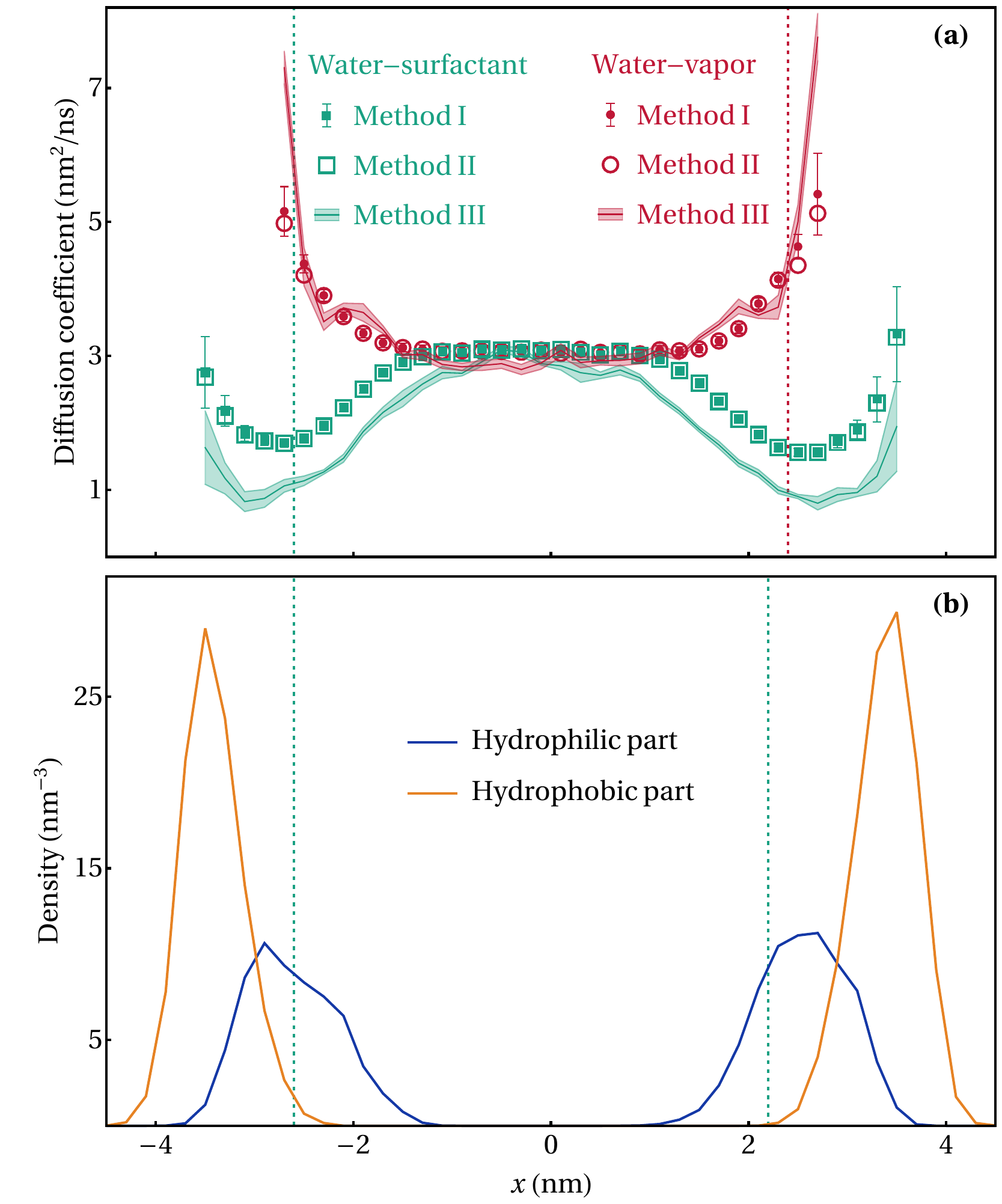}
  \caption{
    \label{fig:main} Estimating the hetereogeneous diffusion coefficient of \ce{H2O} molecules in the water-vapor (red) and water-surfactant (green) systems.
    Panel (a) compares estimations of the transverse diffusion coefficient $D(x)$ as a function of the distance $x$ from the center of the simulation box obtained by three inference methods: (I) the first-passage method [Eq.~\eqref{eq:inferD}], (II) the method of Ref.~\cite{belousov2020}, (III) the method of local mean squared displacements (Appendix~\ref{app:msd}).
    Panel (b) shows the density of hydrophilic and hydrophobic residues of the \ce{C6H12} molecules in the water-surfactant interface.
    Error bars in (a) are given by three standard deviations. Red and green dotted lines indicate the Gibbs dividing interfaces as described in Fig.~\ref{fig:md}.
  }
\end{figure}

Figure~\ref{fig:main}(a) summarizes the results of the three inference methods for $D(x)$ applied to the molecular-dynamics simulations of water-vapor and water-surfactant interfaces. In agreement with the theory of Sec.~\ref{sec:thr} the Fokker-Planck and Smoluchowski approaches (Method I and Method II, respectively) render equivalent estimates of the diffusion coefficient in one-dimensional systems. These values match perfectly the results obtained with Method III in the region of a flat effective potential $U(x)$. Closer to the phase interfaces the performance of Method III deteriorates due to the gradient $\partial_x U(x)$ rapidly changing with $x$.

One may appreciate important qualitative differences between the phase interfaces in
the two simulated systems (Fig.~\ref{fig:main}). The water molecules, which are far away from the surfactant layer, do not experience a gradient of chemical potential. Inside the layer however their motion is restricted
by the surfactant's hydrophilic and hydrophobic groups, which thus decrease the escape rate $\Gamma(x)$
and, consequently, also the diffusion coefficient, \textit{cf.} Eq.~\eqref{eq:Ddef}. The slowdown in the diffusion constant of water is roughly a factor of two relative to the bulk diffusion. Interestingly, we also observe that there is a minimum in the diffusion constant within the surfactant layer which correlates with the transition from the hydrophilic to hydrophobic part of the surfactant shown in Fig.~\ref{fig:main}(d).
 
In contrast, the unsaturated chemical interactions at the liquid-vapor interface
increase both the effective potential energy of water molecules and their escape
rate. In this case, the diffusion coefficient of water molecules near the phase boundary
is therefore larger than in the bulk in agreement with previous studies~\cite{liu2004,liu2005}.
These results suggest that the effective viscosity of water at the interface with vapor
is smaller than near the surfactant film and also varies within the surfactant layer. This type
of microscopic information will provide valuable input for continuum hydrodynamic models of gas diffusion through
phase interfaces~\cite{falciani2020}.

\section{Conclusion}
In this paper we have thoroughly studied the statistical physics of hetereogeneous diffusion  at  macroscopic, mesoscopic, and microscopic scales, both in equilibrium and nonequilibrium conditions. We have shown that Smoluchowski and Fokker-Planck diffusion laws are equivalent and can be used interchangeably, yet the latter is beneficial for inference of space-dependent diffusion coefficients. In particular, we have found that the Fokker-Planck potential and the diffusion coefficient can be extracted by collecting statistics of first-passage times from stochastic trajectories, without  measuring local densities, see Eq.~\eqref{eq:inferD}. We have verified extensively this technique and explored its theoretical implications with applications to soft matter, as discussed below.

As shown in Sec.~\ref{sec:thr} the Smoluchowski equation and Fick's law of diffusion
incorporate the classical theory of statistical mechanics and thermodynamics. In
equilibrium this formulation of the transport problem reproduces the Maxwell-Boltzmann
statistics of mass density. Using Onsager's approach we showed that nonequilibrium
systems are also consistent with the Smoluchowski picture. Furthermore this theory
entails nontrivial relations between the components of a diffusion tensor (Appendix~\ref{app:DT}).

Although the mass transport problem can be formulated in a physically equivalent
form as the Fokker-Planck equation, the associated potential \rb{term must be recognized as a distinct quantity---}not a standard thermodynamic one. As a physical variable the Fokker-Planck potential \rb{belongs
to the microscopic and mesoscopic description of the system: it determines a preferred
direction of the particle flow. This} interpretation has been revealed in our analysis of the simplified
microscopic dynamics and molecules' first-passage statistics.

In addition, we have pointed out that the Ito-Stratonovich dilemma is resolved by providing a physical interpretation of the potential term. Like the diffusion equations~\eqref{eq:SE} and \eqref{eq:DE}, the Ito and Stratonovich conventions are physically equivalent when interpreted consistently with the theory of thermodynamics.
It prescribes to use in the Langevin dynamics formulated with the Ito calculus the Fokker-Planck potential term, which must be mixed with the thermodynamic potential in the other conventions.

The Onsager's approach is also applicable to thermodynamic processes other than
the mass transport. In fact our simplified model of the stochastic kinetics and the
ensuing formalism of stochastic thermodynamics are quite general (Sec.~\ref{sec:mic}).
Combined with the theory of random walks these ideas lead to a consistent interpretation
of the macroscopic thermodynamics and transport equations. The first-passage approach can be formulated for integrated currents of heat, charge, etc. For instance one could measure the time that a given amount of these quantities transferred through a given point.

As a test of our theory, we have developed a robust method to estimate the space-dependent diffusivity of molecules at the nanoscale solely from escape-time statistics, see Eq.~\eqref{eq:inferD}. 
Using this method, we extracted and examined the space-dependent diffusion coefficient in a computational model of a B\"uttikker-Landauer ratchet and in two realistic models of soft-matter interfaces. We have examined the first passage statistics of water molecules at the water-vapor interface and within a surfactant monolayer. Consistently with previous studies, we find that there is a monotonic increase in the diffusion constant as the molecule approach the liquid-vapor phase boundary, whereas within the surfactant the diffusion constant appears to be modulated by the transition from hydrophilic to hydrophobic parts of the surfactant. It will be interesting in the future to explore whether one can use the spatial modulation of first-passage times to detect hydrophillic or hydrophobic "pockets" in living matter.

The asymmetry of soft-matter interfaces with aqueous solutions generate electric fields that could be harnessed for chemical reactions~\cite{Shaik2016,Stefan2017,hassanali2021} relevant in pre-biotic chemistry~\cite{Griffith15697} as well as artificial photosynthesis. Both the statics and dynamics of water at the interface play a crucial role in controlling the magnitude and fluctuations of the electric field. Understanding the changes in water molecules' dynamics, as probed by the space-dependent diffusion coefficient, in such systems may provide important insights for the future models. Our method might find promising applications in examining  such complex geometries and biological environments.

\vfill

\acknowledgments{The authors are grateful to Ricardo Franklin Mergarejo for help in computations. We also thank Daniel Madulu Shadrack, Antonio Celani, Nawaz Qaisrani, and Ali Rajabpour   for stimulating discussions.}

\onecolumngrid
\appendix
\begin{center}\vspace{2em}\textbf{APPENDICES}\end{center}
\section{\label{app:FP2D} Stochastic kinetics in two dimensions}
Equation~\eqref{eq:occupation} in two dimensions can be constructed by considering
a stochastic kinetics with independent transition probabilities along coordinates
$x_1$ and $x_2$:
\begin{eqnarray}
  P&&_{00}(x_1,x_2) = P[(x_1, x_2), (x_1,x_2)]
    = [1-\Gamma_1(x)dt] [1-\Gamma_2(x)dt],\\
  P&&_{0+}(x_1,x_2) = P[(x_1, x_2), (x_1,x_2+dx_2)]
    = [1-\Gamma_1(x)dt] \Gamma_2(x)dt \frac{\me^{\mathcal{L}(x_1,x_2+dx_2)}}{Z(x_1,x_2)},\\
  &&...\nonumber\\
  P&&_{++}(x_1,x_2) = P[(x_1, x_2), (x_1+dx_1,x_2+dx_2)]
    = \Gamma_1(x)\Gamma_2(x)dt^2 \frac{\me^{\mathcal{L}(x_1+dx_1,x_2+dx_2)}}{Z(x_1,x_2)},
\end{eqnarray}
which yield
\begin{multline}
  \partial_t \rho = 
    - \partial_1 [\rho (P_{+0} - P_{-0})] \frac{dx_1}{dt}
    + \partial_{11}[\rho (P_{+0} + P_{-0})] \frac{dx_1^2}{2 dt}
    - \partial_2 [\rho (P_{0+} - P_{0-})] \frac{dx_2}{dt}\\
    + \partial_{22}[\rho (P_{0+} + P_{0+})] \frac{dx_2^2}{2 dt}
    - \partial_1 [\rho (P_{++} + P_{+-} - P_{-+} - P_{--})] \frac{dx_1}{dt}\\
    - \partial_2 [\rho (P_{++} + P_{-+} - P_{+-} - P_{--})] \frac{dx_2}{dt}
    + \partial_{12}[\rho (P_{++} + P_{--} - P_{-+} - P_{+-})] \frac{dx_1 dx_2}{dt}\\
    + \partial_{11}[\rho (P_{++} + P_{-+} + P_{+-} + P_{--})] \frac{dx_1^2}{2 dt}
    + \partial_{22}[\rho (P_{++} + P_{-+} + P_{+-} + P_{--})] \frac{dx_2^2}{2 dt}\\
    =
      - \partial_1 \left\{
        \partial_1 \mathcal{L} \frac{\Gamma_1 dx_1}{dt} \rho
        - \partial_{1}\left[ \Gamma_1 \frac{dx_1^2}{2 dt} \rho \right]
      \right\}
      - \partial_2 \left\{
        \partial_2 \mathcal{L} \frac{\Gamma_2 dx_2}{dt} \rho
        - \partial_{2}\left[ \Gamma_2 \frac{dx_2^2}{2 dt} \rho \right]
      \right\}.
\end{multline}
By identifying in the above equation the Fokker-Planck potential
$$
  \tilde{U} = -2 k_B T \mathcal{L}
$$
and the components of a diagonal diffusion tensor given by Eq.~\eqref{eq:Ds}, we
obtain Eq.~\eqref{eq:DE} in two dimensions.

\section{\label{app:extra} Stochastic kinetics out of equilibrium}
The simplified microscopic dynamics, formulated for equilibrium systems in Sec.~\ref{sec:mic} can be generalized to the nonequilibrium case. To do so we introduce into Eq.~\eqref{eq:moves} a factor, which cannot be derived from the directing function $\mathcal{L}(x)$:
$$
    \frac{P_\pm(x)}{1-P_0(x)} \propto
        \me^{\mathcal{L}(x\pm dx) \pm \epsilon \phi(x)/2},
$$
in which the parameter $\epsilon$ controls the deviation from equilibrium due to the field $\phi(x)$, cf. Sec.~\ref{sec:mac}. The above formula
leaves invariant the normalization factor $Z(x)$, but modifies the transition rates
$$
    k_\pm(x) \propto \me^{-Q(x)dt \pm \partial_x \mathcal{L}(x) dx \pm \epsilon \phi_x/2}.
$$

With the above definitions a new term appears in Eq.~\eqref{eq:occupation}:
$$
    \partial_t \rho = \partial_x \left\{
    - \Gamma dx^2 \rho \left(
        \partial_x \mathcal{L} + \epsilon \frac{\phi}{2}
    \right) + \partial_x \left[\frac{\Gamma dx^2}{2} \rho\right]
  \right\},
$$
which leads to an extra force term $\partial_\epsilon F = D(x) \phi(x)$
in the resulting Fokker-Planck equation,
cf. Eq.~\eqref{eq:gen} and \eqref{eq:Fe}. This formalism may also be extended in a straightforward manner to a vector of nonequilibrium parameters $\bm{\epsilon}$ treated in Sec.~\ref{sec:mac}.

One-dimensional nonequilibrium systems imply however a strong constraint on the form of $\phi(x)$, which we can establish through the Kirchhoff rule for the steady-state current:
\begin{equation}
    0 = \rho(x) P_+(x) \\- \rho(x+dx)P_-(x+dx)
    + \rho(x-dx) P_+(x-dx) \\ - \rho(x) P_-(x)
    \simeq \epsilon \partial_x \phi(x) dx.
\end{equation}
Hence in one dimension we must observe $\phi(x) \equiv \phi_0 = \const$. In presence of more degrees of freedom the Kirchhoff rule contains additional terms, which allow for more general forms of $\phi(x)$.

As we noted in Sec.~\ref{sec:thr}, the one-dimensional systems are always ``conservative'' \cite{lau2007}, in the sense that we always may find a potential form
    $U_\phi(x) = \epsilon k_B T \phi_0 x$,
such that $\phi(x) \propto -\partial_x U_\phi(x)$. However in certain circumstances it cannot be interpreted as the system's thermodynamic energy. One example is the diffusion on a ring, like the B\"uttiker-Landauer ratchet considered in Sec.~\ref{sub:ld}. The system's potential energy must be consistent with the macroscopic symmetry---the periodicity in this example. While a constant function $\phi(x) \equiv \phi_0$ satisfies this requirement, the ``pseudo''-potential $U_\phi(x)$ is not periodic. Hence there exists no potential that would be compatible with the system's symmetry and generate a constant force.

The potential of the Büttiker-Landauer ratchet (Sec.~\ref{sub:ld}) is periodic and, therefore, consistent with the system's symmetry. Consequently it generates no macroscopic current in equilibrium. However, by adding a constant force term to the B\"uttiker-Landauer ratchet, one can generate a nonequilibrium steady-state with a constant current.

Another example is diffusion on a real line, with a bounded potential $U(x)$. The ``unwrapped'' B\"uttiker-Landauer ratchet also fits such a description. In this interepretation the Boltzmann factor of the form $\me^{-[U(x)-U_\phi(x)]/(k_B T)}$ has a minimum at infinity. Therefore the system can never reach a steady state.
Any other function $\phi(x)$, which vanishes at infinity, can be incorporated into the potential energy.

Finally, we remark that the change of the environment entropy
\begin{equation}\label{eq:ness}
    s(x,x+dx) = k_B \partial_x \left[2 \mathcal{L}(x) + Q(x) dt\right] dx
        + \frac{\epsilon}{2} \left[\phi(x+dx) + \phi(x)\right]
\end{equation}
cannot be in general interpreted as the change of the nonequilibrium Boltzmann entropy
$$ S(x) = k_B \ln \rho(x) $$
of the steady-state density $p(x)$, because, as discussed above, the term
$$
    \frac{\epsilon}{2}\left[\phi(x+dx) + \phi(x)\right] \simeq
        \epsilon \int_x^{x+dx} dy \phi(y)
$$
in general is not associated with the total differential of a macroscopic thermodynamic function.

\section{\label{app:DT} Diffusion tensor}
\rb{In principle a nondiagonal diffusion tensor $\mathsf{D}$, whose components always must form a symmetric positive-definite matrix, can be brought
to a diagonal form $\bar{\mathsf{D}}_{ij} = R_{ik} R_{jl} \mathsf{D}_{kl}$ by an
orthonormal matrix $R_{ij}$. This matrix, which encodes a rotation of the coordinate
axes, is parametrized by three rotational angles $\phi_{i=1,2,3}$ in three dimensions
and by one rotational angle $\phi$ in two dimensions. Therefore in general Eq.~\eqref{eq:Q}
and \eqref{eq:ddQ} establish relations between six variables in three dimensions
[three diagonal components $D_{i=1,2,3}(\bm{x})$ and three angles $\phi_{i=1,2,3}(\bm{x})$] and between three variables in two dimensions [two diagonal components $D_{i=1,2}(\bm{x})$ and one angle $\phi(\bm{x})$].}

\rb{For example, in two dimension a general diffusion tensor may have the form
$$
  \mathsf{D}(\bm{x}) = \begin{pmatrix}
    \cos^2\phi(\bm{x}) D_1(\bm{x}) + \sin^2\phi(\bm{x}) D_2(\bm{x}) &
    \cos\phi(\bm{x}) \sin\phi(\bm{x}) [D_2(\bm{x}) - D_1(\bm{x})] \\
    \cos\phi(\bm{x}) \sin\phi(\bm{x}) [D_2(\bm{x}) - D_1(\bm{x})] &
    \cos^2\phi(\bm{x}) D_2(\bm{x}) + \sin^2\phi(\bm{x}) D_1(\bm{x})
  \end{pmatrix}.
$$
Substituted into Eq.~\eqref{eq:Q} the above expression leads to a formidable equation for $h(x_1, x_2)$. Under some geometric constraints these equations may however become tractable.}

Situations, in which the off-diagonal components of a diffusion tensor must vanish
to comply with the macroscopic symmetry of a physical system, \rb{are} quite common.
In fact the two examples considered in Sec.~\ref{sub:md} fall into this category.
Both systems studied there are invariant with respect to rotations about the $x$-axis,
which is transverse to the phase interfaces in the slab geometry. In this case we
can find a general form of the function $h(\bm{x})$ that must satisfy Eqs.~\eqref{eq:Q} and \eqref{eq:ddQ}.

Below we consider Eqs.~\eqref{eq:SE} and \eqref{eq:DE} with a diagonal diffusion
tensor $\mathsf{D}(\bm{x})\to\bar{\mathsf{D}}(\bm{x})$ in two dimensions:
$$
  \bar{\mathsf{D}}(\bm{x}) = \begin{pmatrix}
    D_1(\bm{x}) & 0 \\ 0 & D_2(\bm{x})
  \end{pmatrix}
$$
with positive space-dependent components $D_1(x_1,x_2)$ and $D_2(x_1,x_2)$. Equations~\eqref{eq:Q}
and \eqref{eq:ddQ} then yield
\begin{eqnarray}\label{eq:Urel}
  &&U(\bm{x}) = \tilde{U}(\bm{x}) + k_B T \ln D_1(\bm{x}) - k_B T \phi_1(x_2)
    = \tilde{U}(\bm{x}) + k_B T \ln D_2(\bm{x}) - k_B T \phi_2(x_1),\\
  &&\partial_1\partial_2 h(\bm{x}) = \partial_1\partial_2 \ln D_1(\bm{x})
    = \partial_1\partial_2 \ln D_2(\bm{x}),
\end{eqnarray}
in which $\phi_1(x_2)$ and $\phi_2(x_1)$ are arbitrary real functions of the single coordinates
$x_1$ and $x_2$. The above relations restrict admissible forms that the spatial dependency
of the diffusion tensor may assume:
\begin{equation}\label{eq:D1D2}
  \bar{\mathsf{D}}(\bm{x}) = \rb{d}(x_1,x_2) \begin{pmatrix}
    \me^{\phi_1(x_2)} & 0 \\ 0 & \me^{\phi_2(x_1)}
  \end{pmatrix}
\end{equation}
with the common factor function $h(\bm{x}) = \rb{d}(x_1,x_2)$. \rb{Generalization of Eq.~\eqref{eq:D1D2} for a diagonal diffusion tensor in three dimensions is straightforward.}

\section{\label{app:md} Computational details}

Classical molecular-dynamics simulations of the water-vapor and water-surfactant systems (Sec.~\ref{sub:md}) were performed using the GROMACS package~\cite{abraham2015}. For the surfactant we employed a force field previously validated by Striolo and co-workers, details of which can be found in Ref.~\cite{Shi_2010}. United atoms are used for the alkyl groups, whereas the ester and hydroxyl \ce{OH} groups are treated explicitly. The bonded and non-bonded interaction potentials were modeled by a combination of TRaPPE-UA~\cite{Martin1999} and OPLS~\cite{Briggs1990,Jorgensen1986} force fields. The surfactant component consisted of 96 \ce{C6H12} molecules, 48 on each side of the simulation box. Both simulated systems included 4055 water molecules described by the SPC potential~\cite{Berendsen1987}, which has been shown to accurately reproduce thermodynamic properties, such as the second virial coefficient and the surface tension of the air-water interface~\cite{Zhang2011}.

Periodic boundary conditions are applied in all directions. In water-surfactant simulations a vacuum region of~\SI{215}{\angstrom} in width was reserved between the two monolayers of \ce{C6H12} molecules. A cutoff of \SI{1}{nm} is used in the real space for the Coulomb interactions treated by the full-particle mesh Ewald method, as well as for the Lennard-Jones interactions. All simulation models were first equillibrated in the NVT-ensemble for \SI{10}{ns} using the  Nose-Hoover thermostat~\cite{Nose1984,Hoover1985} at \SI{298}{K} with a relaxation constant of \SI{2}{ps}. For the first passage analysis we produced 0.5-\si{ns}-long trajectories with the positions of the particles saved every \SI{10}{fs}.

\section{\label{app:msd} Local mean squared displacements}

The method of mean squared displacements, which is commonly used to estimate
the diffusion constant in homogeneous systems, relies on the Einstein formula for
the mean squared displacement in absence of the potential $U(x)$:
\begin{equation} \avg{[x(t)-x(0)]^2} = 2 D t. \end{equation}
The above formula is not applicable to systems with an external potential or with
the diffusion coefficient, which varies along the axis of displacements. Therefore
the standard method of mean squared displacements is not suitable for measuring the
transverse diffusion coefficient~\cite{belousov2020}.

For a Brownian particle however another formula holds regardless of the external
potential, which may be present in the system \cite{lamouroux2009,baldovin2018,baldovin2019}:
\begin{equation}\label{eq:lim}
  D[x(0)] = \lim_{t\to0} \frac{\avg{[x(t)-x(0)]^2}}{2  t}.
\end{equation}
In a real physical system or a molecular-dynamics simulation Eq.~\eqref{eq:lim}
is not strictly valid: due to the molecules' inertia \cite{belousov2016II} there exists a ballistic regime of diffusion at small times $t$ with $\avg{[x(t) - x(0)]^2}\propto t^2$, \textit{cf.} Chapter II in \cite{chandrasekhar1943}.

Interpreting Eq.~\eqref{eq:lim} in a physical sense \cite{baldovin2018,baldovin2019},
we may choose a sufficiently small time $t$ (larger than the momentum relaxation time) and use an approximate expression
\begin{equation}\label{eq:apx}
  D(x) \approx \frac{\avg{[x(t)-x(0)]^2}}{2 t},
\end{equation}
conditioned on $x(0) \in [x_i-L,x_i+L]$,
in which the integer $i$ enumerates the regions of interest as in the first-passage method (Sec.~\ref{sec:app}). Then we choose a small interval
similar in order to the mean first-passage time $\tau$. Over this interval the molecules'
displacements remain on average within the scale of $L$, though a small fraction of observations would probe properties of the neighbor regions.

As follows from the above description, we obtain mean squared displacements for molecules
exploring the local environment in a given region of interest. These data can be fitted by a
line to extract the spatially resolved diffusion coefficient $D(x)$ as in the standard
method of mean squared displacements (Fig.~\ref{fig:msd}). Although in our approach
we rely on Eq.~\eqref{eq:apx} and thus do not account for the external field $U(x)$,
a correction to the quadratic or higher orders in time---for the gradient and the curvature
of the potential respectively---could be in principle derived. Due to this fact,
the described method works most reliably when the diffusion coefficient is small and the potential $U(x)$ is flat.

\twocolumngrid

\bibliographystyle{apsrev4-2}
\bibliography{main}
\end{document}